# Guidelines in Wastewater-based Epidemiology of SARS-CoV-2 with Diagnosis

**Madiha Fatima** [1][†], **Zhihua Cao**[2][†], **Aichun Huang**[2][†], **Shengyuan Wu**[2], **Xinxian Fan**[2], **Yi Wang**[2], **Liu Jiren**[5], **Ziyun Zhu**[2], **Qiongrou Ye**[2], **Yuan Ma**[2], **Joseph K.F Chow**[3], **Peng Jia**[2], **Yangshou Liu**[2], **Yubin Lin**[2], **Manjun Ye**[2], **Tong Wu**[2], **Zhixun Li**[2], **Cong Cai**[2], **Wenhai Zhang**[2], **Cheris H.Q. Ding**[4]*, **Yuanzhe Cai** [2]*, **Feijuan Huang** [5]*

1. Shenzhen Second People's Hospital, Shenzhen WEIMEI Medical Technology Co., Ltd. Luhou District, Shenzhen, Guangdong, China; madiha.fatima@wus.edu.pk
2. College of Big Data and Internet, Shenzhen Technology University, 3002 Lantian Road, Pingshan District, Shenzhen 518118, China; caiyuanzhe@sztu.edu.cn; 564849152@qq.com; huangaichun04@gmail.com; 1511731576@qq.com; 1950768640@qq.com; wang2954295926@163.com; liujieren2018@stu.hnucm.edu.cn; 3318931792@qq.com; yeqiongrou@163.com; 2993494019@qq.com
3. ThunderBio Innovation Limite, Skyworth Innovation Valley, No. 8 Tangtou No. 1 Road, Baoan District, Shenzhen, Guangdong, China; kfcjoseph@gmail.com
4. School of Data Science, Chinese University of Hongkong, No 2001 Longxiang Boulevard, Longcheng Street, Longgang District, Shenzhen 518172, China; ChrisDing@cuhk.edu.cn
5. Shenzhen Second People's Hospital, The First Affiliated Hospital of Shenzhen University, Shenzhen, China; huangfeijuan@163.com

\* Correspondence: huangfeijuan@163.com; caiyuanzhe@sztu.edu.cn; Cheris Ding@cuhk.edu.cn

† These authors contributed equally to this work.

**Abstract:** With the global spread and increasing transmission rate of SARS-CoV-2, more and more laboratories and researchers are turning their attention to wastewater-based epidemiology (WBE), hoping it can become an effective tool for large-scale testing and provide more accurate predictions of the number of infected individuals. Based on the cases of sewage sampling and testing in some regions such as Hong Kong, Brazil, and the United States, the feasibility of detecting the novel coronavirus in sewage is extremely high. This study reviews domestic and international achievements in detecting SARS-CoV-2 through WBE and summarizes four aspects of COVID-19, including sampling methods, virus decay rate calculation, standardized population coverage of the watershed, algorithm prediction, and provides ideas for combining field modeling with epidemic prevention and control. Moreover, we highlighted some diagnostic techniques for detection of the virus from sewage sample. Our review is a new approach in identification of the research gaps in waste water-based epidemiology and diagnosis and we also predict the future prospect of our analysis.

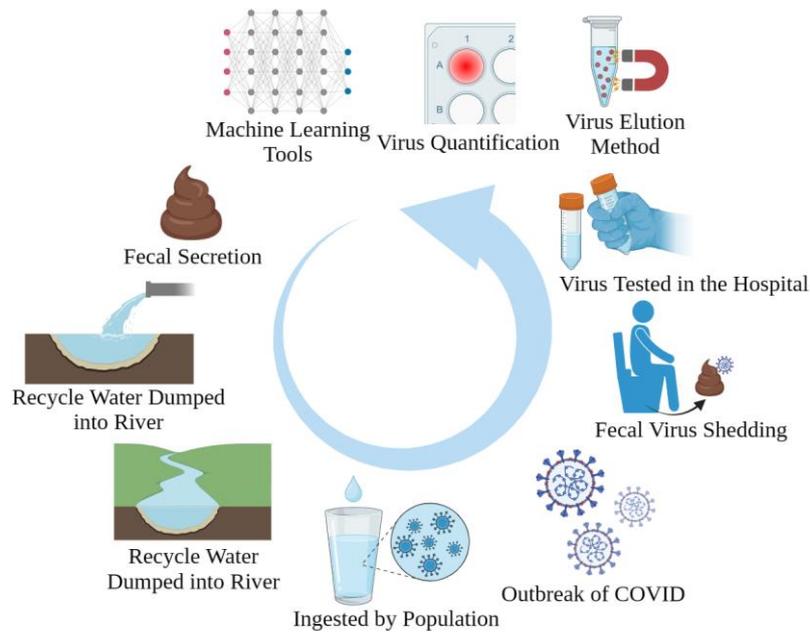

**Keywords:** Novel coronavirus pneumonia; Wastewater-based epidemiology (WBE); Machine learning

1. **Introduction**

   In recent years, with the global spread of SARS-CoV-2, wastewater-based epidemiology (WBE) has been used as an important strategy to detect pathogens excreted by infected individuals and to estimate the spread of the virus [1-2]. SARS-CoV-2 RNA concentrations in wastewater have been shown to be an early indicator of rising COVID-19 community incidence. Emerging variants of concern (VOCs) and variants of interest (VOIs) demonstrate increased transmissibility, disease severity or immune escape [3]. Timely and accurate quantification of local prevalence of SARS-CoV-2 variants is thus essential for effective public health measures [3,4].

   Due to the relatively low cost of WBE detection [4], it can effectively alleviate the economic burden of local COVID-19 testing. As a public health surveillance method, WBE has a huge advantage over traditional and clinical surveillance in an epidemic such as SARS-CoV-2 because it uses a pooled sample of the community and can generate comprehensive population data without relying on test individuals and reporting test results in traditional clinical surveillance [5].

   Wastewater-based epidemiology (WBE) has emerged as a potential tool for tracking COVID-19 infection rates in surrounding areas of collection points. This method uses grab, composite and passive samplers, and machine or deep learning algorithms that can be applied to process environmental and biological factors for more timely and accurate predictions of infection numbers [6–12]. Since the outbreak of the COVID-19, artificial intelligence tools have been used in tracing origin of the infection, diagnosis, and preventive control strategies against the virus [4]. The machine learning models such as time-series, non-time-series models and long short-term memory (LSTM) models have used to trace viral (SARS-CoV-2) loads and location identity via WBE [13].

In order to effectively monitor the spread of viruses, sewage testing has become a very effective method. Based on practical cases in regions such as Hong Kong, Brazil, and the United States, it has been found that detecting the novel coronavirus in sewage is highly feasible [6-8]. However, there are still many uncertainties that interfere with the prediction's outcomes, such as the pH value of sewage, precise measurement of sewage temperature, and cross-infection events are difficult to avoid during sewage sampling. Limited experiments in this field present challenges that may interfere with predictions. Therefore, addressing these challenges is a key research direction in realizing the full potential of WBE in the future and highlights the significance of exploring this method in the ongoing pandemic.

In this review, we summarize the recent progress and future directions in the field of WBE and propose an integrated virus sampling, enrichment, elution, and detection technology to facilitate effective monitoring of SARS-CoV-2 in wastewater.

## 2. The impact of sampling method selection on the results of wastewater-based epidemiology

### 2.1 Sample collection

Since the global spread of SARS-CoV-2, wastewater-based epidemiology (WBE) has been used for detecting and estimating the spread of the virus. WBE samples generally consist of grab samples and composite samples, and the use of passive samplers is also gradually emerging as collection of waste methods. The detailed sampling methodology and mechanism is described in Table 1 and also explained with limitation in below section [14-15, 17].

#### 2.1.1 Grab Sampling

The advantage of grab sampling is that it can quickly collect samples within a short period of time without the need for deployment of automated equipments at sampling points. Additionally, studies have shown that grab sampling during the peak fecal load period can effectively increase the target virus concentration in the sample. Therefore, the peak fecal load time should be determined before sampling [16]. However, the limitation is that the discrete samples obtained by grab sampling can only represent the sewage condition at the time of collection, and fluctuations in flow rate and composition of the sewerage system at that time also have significant effects on sample detection. Additionally, studies have shown that grab sampling during the peak fecal load period can effectively increase the target virus concentration in the sample. Therefore, the peak fecal load time should be determined before sampling [18].

#### 2.1.2 Composite samples

Composite samples collect a large number of samples at fixed time intervals and the samples are subsequently combined in proportion to the effluent flow rate. This sampling mode provides a well-mixed sample that is more representative [19,20], and which represents the average effluent characteristics over the collection period. According to the study, composite samples eliminate peak inconsistencies and are more reliable than grab samples; while grab sampling techniques are more convenient for sampling in the face of unstable constituents that require timely analysis [21].

### 2.1.3 Passive samplers

| Sampling method | Collection mechanism | Advantage | Disadvantage |
|---|---|---|---|
| **Grab sampling** | Through certain sampling equipment, some substances in sewage are taken out as samples for testing or analysis. Its main principle is to use the equipment to collect and maintain the fluid function, will meet the sampling requirements of the sewage sample out, and minimize the influence of external intervention in the sampling process. | Samples can be collected quickly in a short period of time without the need to deploy automated equipment at the sampling point; Sampling during peak fecal load can effectively increase the concentration of target virus in the sample. | The discrete samples obtained by capturing samples can only represent the sewage status at the time of collection, and the fluctuations of the flow and composition of the sewage system at that time also have a significant impact on the sample detection |
| **Composite sample** | A large number of samples are collected at fixed intervals and then combined in proportion to the outflow rate. | This sampling mode provides a well-mixed sample, which is more representative; Composite samples eliminate peak inconsistencies and are more reliable than grasping samples | As multiple samples are mixed together, it is not possible to obtain the detailed information that a single sample can provide, and it can also lead to changes in the sample, such as an increase in the number of microorganisms or degradation of contaminants, which can affect the accuracy of the test results. |
| **Passive sampler** | A passive sewage sampler is a device used for sewage sampling, usually consisting of a series of pipes and containers. The principle is to guide the sewage to be measured into the sampler for collection through natural flow. | It can be deployed at a specific location in a wastewater system for a specific period of time, and its deployment is quick, simple and inexpensive without the need to enter a confined space, but is more complex than grabbing samples from a sampling quantitative infection level. | Sampling from sampling quantitative infection levels is more complex. |

Passive samplers can be deployed for a specific period of time at a specific location in the wastewater system, and their deployment is rapid, simple and inexpensive, without the need to enter confined spaces, but more complex compared to grab sampling from sampling quantitative infection levels aspect.

A practical, convenient and inexpensive molar swab method has also been developed for the collection and analysis of sewage samples. The collected molar swab samples are more sensitive than grab samples and are simple and inexpensive[16,17].

### 2.2 Sampling frequency

Table 1. Sampling methodology with mechanism of waste water sample

Selecting the frequency of sampling in a purpose-oriented manner, if the goal of effluent monitoring is to screen effluent for the presence of SARS-Cov-2, therefore, a sampling frequency of once a week is sufficient. If an early indication of infection trends is desired, monitoring with at least three sampling sites over the target trend period needs to be deployed [22,23]. One study proposed a random forest prediction performance model that illustrates the variability of RF performance within the same region. The prediction results of this study show that sampling frequency is a key parameter for improving RF performance; the results show that sampling more than once a week is necessary to produce reasonable predictions, and the greater the frequency, the higher the accuracy [24].

## 2.2 Sampling locations

According to a survey of relevant institutions, there is no publicly complete framework of sampling site selection methods that can maximize the detection value of RNA effluent detection of SARS-COV-2 in communities. However, Yaeger has proposed a practical location-based method to select raw sewage sample locations, which effectively establishes epidemiological and geolocation-based urban communities for the monitoring of RNA of SARS-COV-2 in sewage [25].

Also, sampling locations can be chosen to include the entire WWTP (wastewater treatment plant) catchment area for community-level monitoring or to target smaller communities and areas for targeted monitoring. Effluent samples collected from wastewater treatment plants can provide an approximate overall picture of disease prevalence at the town-wide level. However, samples with a finer regional breakdown can provide a more detailed picture in disease prevalence, allowing for a more comprehensive analysis and prediction of the data [26]. Detailed hydraulic models of urban sewer networks proposed in some studies can largely help to determine the best sample locations for target monitoring of complex sewer networks (Figure 1) [27,28]. In the case of static surveillance, where the sampling location and number are predetermined and the location remains fixed, the sampling location selection and optimization problem usually ensures maximum coverage of the population. In the case of dynamic surveillance, where the sample location and number are uncertain, this situation can be solved by established sampling location and optimization to detect the specific location of virus presence or locate the additional outbreaks [29].

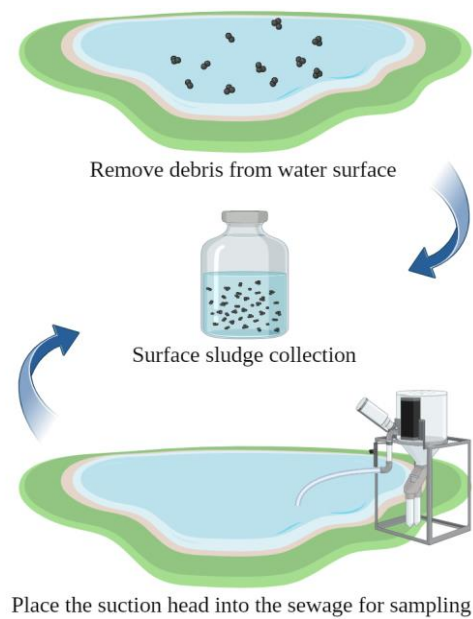

**Figure 1. The figure depicted the sample collection methodology and hydraulic models of urban sewer networks.**

## 3. Enrichment units

### 3.1 Enrichment principle

Wastewater treatment and enrichment methods can be divided into four main categories according to their role: physical, chemical, biological and physio-chemical methods 错误!未找到引用源。.

1. The physical method uses physical action to achieve the purpose of separating suspended pollutants in sewage, it is simple to operate and the physical properties do not change during the treatment process, the main methods include sedimentation and filtration [30].
2. The chemical method involves the addition of chemical reagents to the effluent and the use of chemical reactions to achieve the separation of pollutants or make them harmless. The treatment results are good, but are costly and require pre-treatment [30].
3. The biological method uses natural microorganisms to oxidize and decompose organic pollutants or inorganic toxins dissolved in the sewage, and transform them into harmless inorganic substances to achieve the purpose of purifying the sewage 错误!未找到引用源。.
4. Physio-chemical methods use the basic principles of adsorption, extraction and other methods, membrane separation technology, ion exchange and other technologies to separate inorganic or organic pollutants from wastewater, the main methods include extraction and adsorption, membrane analysis (including dialysis, electrodialysis, reverse osmosis, ultrafiltration, etc.), when the concentration of impurities is very low or high.

Based on the knowledge of wastewater treatment methods, the physical methods include flocculation, precipitation, membrane filtration, elution, ultracentrifugation and immunomagnetic beads, etc., the chemical methods include PEG precipitation, etc., [32] the physio- chemical methods include ultrafiltration, solid particle adsorption and elution, etc. The detail process of sampling, elution and enrichment is explained in Figure. 2 [23, 33].

### 3.2 Elution principle and effect of elution parameter

Elution, re-extraction of the virus from the filter membrane by altering its external electrical properties. At present, the most common elution method is cationic environment, the bioelectricity of viruses that can attack the human body is mostly negative, so the positive filter membrane (ultrafiltration membrane, microfiltration membrane) can be used to intercept and adsorb the negatively charged virus.

### 3.3 Elution parameters

During the elution process, the charged barrier of the virus is damaged by the elution solution and loses electrical property, so it falls off from the filter membrane. In order to increase the elution efficiency, the eluent with high elution efficiency should be selected and the elution environment of the virus should be paid attention. In the process, the lotion should be soaked for several times and rinsed in circulation.

In the evaluation of the effect of different eluents on virus elution in water by Zhang study [24], 3% beef immersion powder and 8% beef extract were used to elute the known amount of virus on the colander, respectively. The first, second and third elution rates were 32.14%, 39.24% and 31.24%, respectively, when 3% beef powder was used for elution. By comparison, the elution rates of 8% beef extract for virus could reach 37.01% and 49.66%, respectively, which were higher than that of 3% beef extract powder. Moreover, after adding positive salt ions (such as

MgCl2), it was concluded that cations could enhance the desorption effect of virus at high PH value. In the study on the influence of PH value on coronavirus, the presence of novel coronavirus was detected in feces through experiments, and in the acid-base study on the survival conditions of SARS virus, it was proved that when the PH value could reach 9, the coronavirus could survive for a longer period of up to 4 days. In the report of Zhang Chuyu's team, it is pointed out that the PH value of the eluent is not limited to 9, but maintaining it near 9 can ensure that the virus does not loss activity. Therefore, the survival rate of the virus and the elution efficiency can be improved by appropriately improving the acidity and alkalinity of the PH value of the water.

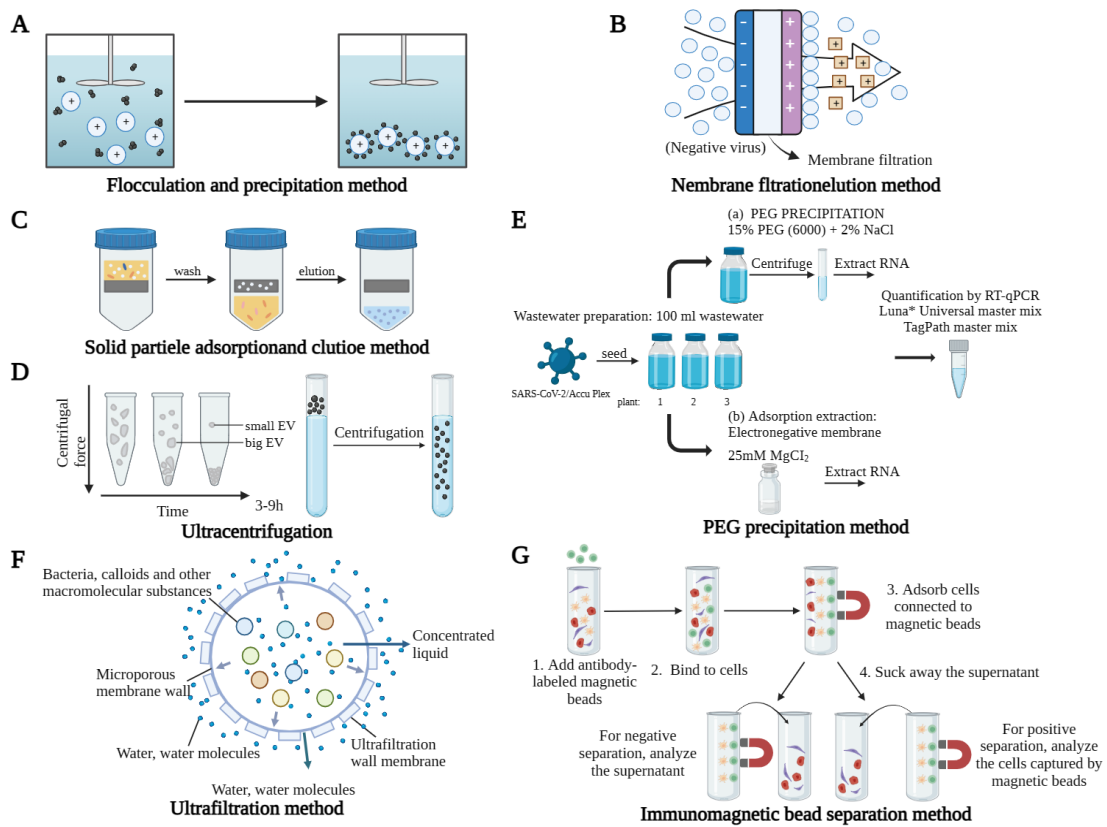

**Figure 2. The schematic figure represents the principle of sampling, elution and enrichment of the samples.**

## 4. Potential markers for SARS-CoV-2 in wastewater epidemiology

It is critical to interpret the viral load by the population biomarkers in wastewater by normalizing SARS-CoV-2 concentration comparing the epidemiological trends among the sewer sheds, and identifying the vulnerable communities [34]. The essential characteristics of biomarkers are to maintain some stability in the wastewater system in the presence of high concentrations. Moreover, excretion needs to have a high level of observability and high susceptibility to biologically infected viruses in sewage. Minimal variation is found in daily excretion within individuals and high biological capacity to identify origins. These are less influenced by dilution of sewage, less interference from exogenous and endogenous factors [35].

PMMoV is one of the most abundant viruses in the human gut and has been identified as one of the molecular targets of fecal indicator viruses in several pepper-based products and diets [36], with a more stable persistence in effluent [25]. An experiment conducted in the Missouri Statewide Wastewater SARS-CoV-2 Detection Program in the United States to study the effectiveness of PMMoV in standardizing SARS-CoV-2 loads. In utilizing a direct normalization approach, PMMoV reduced the correlation between viral load and number of cases. In using the indirect normalization method, PMMoV showed a smaller correlation coefficient between viral load and number of cases, with a large gap with other biomarkers [38]. In addition, experiments on PMMoV and case normalization were done in sewage treatment plants in Hong Kong, and the normalization of PMMoV virus to SARS-CoV-2 viral signal did not strengthen the correlation between the number of clinical cases in sewage treatment plants [39]. However, in Stockholm, a PMMoV-only experiment found a correlation between PMMoV standardization and case volume [40]. PMMoV is influenced by exogenous sources, possibly from chicken and seagull feces, with potential exogenous sources from agricultural soils, suspended sediments and fertilizers. Therefore, it remains to be discussed whether PMMoV is the most appropriate biomarker. Some other biomarkers are HIAA, PARA & CAF and Creatinine. HIAA is one of the main metabolites of serotonin and is less influenced by lifestyle or habit. In the Missouri experiment, the normalized correlation of 5-HIAA was almost the same as PMMoV [38]. Although an endogenous marker that is stable in effluent and meets most effluent epidemiological criteria [41], its potential is also debatable due to its cumbersome extraction processes [38]. PARA is a metabolite produced through human consumption of caffeine-containing products (coffee, tea and caffeinated beverages) and has been recognized as a reliable population biomarker [42]. In the Missouri experiment, [38], making PARA a suitable biomarker with high potential [42]. However, PARA levels were found to be less affected by genetic heterogeneity and population structure compared to its daughter compound PARA in the Missouri experiment, and thus PARA may be a better choice [43]. Creatinine is a metabolite of creatine and phosphocreatine in muscle, but possesses a large amount of sewage diluting creatinine in the sewer system [41,42,44], which affects its potential as a population biomarker. In the Missouri experiment, it had the lowest correlation coefficient and its potential as a marker was small [38].

In comparison of biomarkers to population standardized RNA concentrations of SARS-CoV-2 with COVID-19 incidence, normalization of SARS-CoV-2 RNA concentrations using PMMoV, creatinine, 5-HIAA, PARA and CAF was positively correlated with COVID-19 incidence.

Among them, CAF and its metabolite PARA are less restricted and are potentially the best biomarkers in detecting COVID-19 [38].

There are two methods for standardization, one is direct and the other is indirect. The direct method uses biomarker concentrations and performs calculations to determine the ratio of the population acting on the effluent volume (Pr, population size/L) for direct standardization of the viral load, which is similar to the PMMoV standardization scheme recommended by the CDC and the European Commission; an indirect method, where the population standardization process uses marker names of marker concentrations to replace the estimated population [38].

Biomarkers used for population standardization should have high accuracy and low variability. Some studies have been performed by direct normalization, using normalization factors and biomarker concentrations[39] , so CAF is superior to other biomarkers. The results showed that CAF should be the most suitable biomarker for the direct normalization method, followed by PARA, 5-HIAA, and finally PMMoV [38]. Among the indirect normalization methods, one study pointed out that normalization coefficients were calculated using data-transformed biomarker concentrations, and CRE showed the highest variability and the lowest precision among all biomarkers. Therefore, the most suitable biomarker for the indirect normalization method is PARA, followed by CAF, PMMoV and 5-HIAA [38].

In other studies, it was shown that PMMoV has a role in population normalization, which holds a different opinion on PMMoV internationally [39,40,45]. While 5-HIAA is highly stable [44] , it is difficult to collect and has a low normalization factor, which has some potential, and it is challenging to adopt 5-HIAA as a biomarker [38]. PARA and CAF have a high potential because of their low exogenous effect and high normalization factor.

5. Diagnosis

Nucleic acid and antigen testings get accustomed to sifting potential SARS-CoV-2 carriers out from people at present[46]. Nucleic acid test has high sensitivity and specificity, and is better than antigen detection in clinical trials. According to the recommendation of WHO and the Centers for Disease Control and Prevention (CDC), nucleic acid amplification test (NAAT) by quantitative reverse transcription PCR (RT-qPCR) is regarded as the Normative reference for detecting SARS-CoV-2 错误!未找到引用源。. However, RT-qPCR detection has certain limitations, with the following main limitations: decreased sensitivity of detection results when the virus is at low concentrations, susceptibility to PCR inhibitors during the detection process, false positives in detection results caused by background DNA/RNA contamination, and the need for standard curves to quantify the results [44, 46]. It is worth noted that in patients with clinical imaging suspicion of pneumonia, it may even initially be negative and a few days later be able to detect viral RNA 错误!未找到引用源。. Finally, the sensitivity of RT-qPCR varies depending on the type of sample used 错误!未找到引用源。. Therefore, there is a pressing need for a more precise detection method to detect SARS-CoV-2 and quantify its levels.

In addition to RT-qPCR and other approved diagnostic tests include antigen and antibody testing using LFA or enzyme-linked immunosorbent assay (ELISA) [29]. At present, there are also many fast and efficient new diagnostic techniques, such as clustered regular interspaced short palindrome repeat sequences (CRISPR), next-generation sequencing (NGS) , loop mediated isothermal amplification (LAMP), and ddPCR, which are also available [34, 36]. Digital PCR has various types including chip based or, more commonly, microdroplet based (microdroplet digital PCR; ddPCR) (Figure 3) [34]. Research has shown that ddPCR has the advantages of high sensitivity and accuracy compared to other detection methods, which will play an important role in the diagnosis of fungal, bacterial, and viral pathogens [37]. Studies have shown that ddPCR can accurately detect and quantify SARS-COV-2 even in low Viral load samples. This is because before the amplification steps [39, 47], the initial PCR reaction mixture was divided into multiple reactions by the experiment, and the PCR inhibitor was diluted, which greatly reduces the impact of PCR inhibitor susceptibility. ddPCR is a technique for absolutely quantifying nucleic acid targets by randomly dividing the fluorescence quantifying reaction system into tens of thousands of independent micro-reaction units [48–52]. ddPCR developed from the original PCR to the third generation. Compared with qPCR technology, ddPCR has many advantages: high sensitivity, low detection limit and so on [53–55].

In community settings, ddPCR may be helpful for early detection of SARS-COV-2, especially when individuals at risk of carrying the virus have negative RT-qPCR results, which can greatly improve the accuracy of the test results and help prevent the outbreak of SARS-COV-2 in the community [56]. It is worth noting that there is evidence to suggest the presence of trace amounts of SARS-COV-2 in the saliva of patients with early infection [57], which further validates that

SARS-COV-2 can be detected in sewage and attracts attention in the community. A study conducted by Cassinari et al. showed that the sensitivity of ddPCR is superior to RT qPCR [58].

ddPCR can also play a key role in the disease severity stratification of patients with COVID-19 [59]. In sewage testing, ddPCR detection can not only detect carriers of SARS-COV-2 virus, but also detect the severity of the patient's disease [59, 60], thereby understanding the specific situation of the local SARS-COV-2 outbreak. A research report states that the plasma concentration of RNA in SARS-CoV-2 patients is closely related to the degree of SARS-CoV-2 disease [61]. In addition, higher Viral load in plasma will increase the mortality of patients or the number of ICU admissions [59, 62]. DdPCR is also very accurate in detecting and quantifying RNA plasma levels in patients with immune impairment and ICU patients with SARS-COV-2, which has potential clinical significance [59, 62].

Compared with RT-qPCR, ddPCR has a higher tolerance to PCR inhibitors, which is one of the reasons why ddPCR shows a better detection ability of SARS-CoV-2, which was confirmed in a study by Szwebel et al [63]. The authors reported that ddPCR showed increased tolerance to PCR inhibitors compared to RT-qPCR announced its emerging potential for viral diagnosis [63].

Table 2: Comparison of PCR techniques among different generation

| Intergenerational division | classification | target | characteristic | On behalf of the company | application area |
|---|---|---|---|---|---|
| First generation PCR | First generation PCR | SNP single nucleotide polymorphism site | Quick detection | Xi'an Gold Magnetic Nano | Genetic testing, etc |
| Second generation PCR | Fluorescence quantitative PCR | Pathogenic microbial nucleic acid | High precision, quantifiable | Sun Yat sen University Daan Gene Co., Ltd | HPV, hepatitis virus testing, genetic disease gene testing, etc |
| | Mutation amplification blocking system AEMS | Tumor Mutant Genes | High detection accuracy | Yakangbo | Tumor single gene testing, etc |
| | High resolution dissolution curve HRM | Two site alleles | Can only detect mutations | Wuxi Ruiqi | Tumor single gene testing, etc |
| Third generation PCR | multiplex pcr | Multipoint | Simple, efficient, and affordable | BGI | Simultaneous detection or identification of multiple pathogenic microorganisms, including certain pathogenic microorganisms |
| | digital PCR | Absolute nucleic acid quantification | High sensitivity and expensive price | Thermo Fisher | Research on gene expression differences, etc |

ddPCR can accurately detect and quantify SARS-COV-2 in samples with low viral load, and can correct the limitation of poor sensitivity of RT-qPCR to samples with low viral concentration. Yan Dang et al., collected a total of 117 samples from 30 confirmed patients and 61 uninfected patients. Qualitative and quantitative analysis of these samples were performed by reverse transcriptase quantitative RT-qPCR and ddPCR to evaluate the diagnostic performance and applicability of the two methods. The positive rates of RT-qPCR and ddPCR were 93.3% and 100%, respectively. Three RT-QPCR-negative samples were tested positive by ddPCR. Among 17 low-concentrated virus samples, ddPCR results were all positive, but only 9/17 samples were positive by RT-qPCR. In 2020, Brand et al., found that 8.6% of RT-QPCR-negative results were considered positive by ddPCR in 208 disc-pharyngeal swabs sampled from various hospitals in Greater Paris from March to April. All positive samples detected by reverse transcription RT-qPCR were confirmed by ddPCR and was positive in both samples, while RT-qPCR had high uncertainty. Therefore, compared with RT-qPCR, ddPCR showed higher sensitivity, especially in samples with low viral load 错误!未找到引用源。. The sensitivity of ddPCR sequencing was higher than that of targeted amplicon sequencing (NTS), and the sensitivity of ddPCR sequencing was also higher than that of targeted amplicon sequencing (NTS) in detecting target mutations. 547 wastewater samples (249 in phase I and 298 in Phase II) were analyzed using ddPCR (targeting N1, N2 and 5 mutations) and sequencing. The SARS-CoV-2 wastewater concentration (as determined by the average of N1 and N2 concentrations) was significantly higher in the first stage than in the second stage. The results indicated that the sensitivity of sequencing assay was lower than that of ddPCR assay when focusing on the common targeted N-gene regions in the SARS-CoV-2 genome. Sequencing is also less sensitive than ddPCR in detecting targeted mutations 错误!未找到引用源。. The basic principle of micro droplet ddPCR and chip based ddPCR is to disperse a large amount of diluted nucleic acid solution into the micro reactor in the chip, with each reactor having less than or equal to one nucleic acid template. After the PCR cycle, a reactor with a nucleic acid molecule template will give a fluorescence signal, while a reactor without a template will have no fluorescence signal [35]. Based on the relative ratio and the volume of the reactor, the nucleic acid concentration of the original solution can be calculated. Therefore, ddPCR is compared with other detection methods to highlight the differences in sensitivity and specificity (Table 3).

Comparison of ddPCR, qPCR, and LAMP in the context of SARS-CoV-2 detection [36]. Using EUA approved PCR assay as a reference standard, various test conditions were evaluated, including primer sets, RNA extraction methods, and human specimen types [37]. Overall, ddPCR has the highest diagnostic sensitivity, but its specificity is lower than qPCR and LAMP. Research has found that nucleic acid testing using pharyngeal swabs is superior to saliva testing using Reitsma and latent bivariate models. This finding is consistent with a study conducted from February to April 2021, which compared saliva samples with pharyngeal swabs by qPCR and LAMP detections. Using saliva as starting material, the sensitivity and specificity of detection is only 80-86% and 98-99% respectively, which is close to the summary index (Reitsma model) [38]. The test results also comply with the recommendations of the CDC (Centers for Disease Control and Prevention) in the United States and the World Health Organization regarding the use of throat swabs instead of saliva. Overall, testing using throat swabs should still be the gold standard for diagnosing COVID-19 and a reference standard for rapid detection of SARS-CoV-2 infection in large-scale population screening. In addition, qPCR and LAMP assays applied to samples without RNA extraction have consistently performed poorly in terms of sensitivity and specificity. This finding suggests that compared to tests that include RNA extraction steps, without RNA extraction, it is highly likely that patients infected with SARS-CoV-2 will not be able to be identified. Therefore, methods that do not extract RNA should only be considered when the importance of saving time and clinical resources outweighs the need for high testing performance [39].

When testing for differences in Area Under Curve (AUC) between subgroups, our unpublished data suggested that sample type is a key factor determining AUC in ddPCR and LAMP, while the selection of RNA extraction methods also strongly affects qPCR performance. These observations indicate that specimen type and RNA extraction methods should be the main focus for optimizing these assays. When we limited our analysis to studies using only throat swabs and RNA extraction, the three nucleic acid tests using ORF 1ab primers consistently outperformed those using N primers. This result may be due to fewer mutations in ORF 1ab than in the N region [41].

Importantly, when controlling the experimental settings for each test by specifying primer probe sets, RNA extraction, and sample types, the accuracy of AUC estimation did not differ among different nucleic acid tests. Significant changes in accuracy were observed when the experimental settings were changed. This means that optimization work should focus on designing robust and consistent experimental devices, as the accuracy of SARS-CoV-2 nucleic acid detection depends more on the experimental device rather than the type of detection used. This conclusion also suggests that LAMP and ddPCR may be reliable substitutes for the current qPCR gold standard, as the three nucleic acid tests have similar accuracy when experimental conditions are similar. It is worth noting that using LAMP as a reference standard with similar accuracy as PCR methods will benefit clinical systems or nursing point environments with scarce medical resources in low - and middle-income countries. In high-income countries where ddPCR is more readily available, the advantage of this method is that it provides an absolute number of virus copies in the sample. Therefore, due to its high sensitivity and repeatability, it can be used to calibrate other detection methods. Therefore, different test types can be selected based on environmental and socio-economic requirements, but the test results should be consistent, provided that the best experimental settings are implemented and standardized [42].

In sensitivity analysis of nucleic acid testing, the low detection limit of ddPCR is consistent with its high mixed sensitivity from clinical samples. LAMP research has shown that the maximum change in detection limit distribution may be due to inherent differences in detection methods such as colorimetry and turbidity [43].

With the increasing number of COVID-19 cases and deaths worldwide, there is an urgent need to investigate the effectiveness of current diagnostic tests and the optimal conditions for these tests in order to achieve optimal accuracy and consistency. By summarizing the estimated values of ddPCR, qPCR, and LAMP performance (Table 4), it can be concluded that all three nucleic acid tests consistently outperform saliva samples in pharyngeal swabs and ORF 1ab compared to the N primer probe group. RNA extraction is a key step in ensuring the accuracy of all three tests. ddPCR has been proven to be the most sensitive, followed by qPCR and LAMP. However, there was no significant difference in their accuracy; On the contrary, accuracy depends on specific experimental conditions, which means that more efforts should be made to optimize the experimental settings for nucleic acid testing. Our results can serve as a reference for optimizing and establishing a standard nucleic acid testing protocol suitable for laboratories [44].

Table 3. Overall sensitivity and specificity of LAMP, dPCR, and qPCR

|  | LAMP | dPCR | qPCR |
|---|---|---|---|
| sensitivity | | | |
| Reitsma | 83.3%（76.9-88.2） | 94.1%（88.9-96.9） | 92.7%（88.3-95.6） |
| Latent | 86.2%（20.7-99.9） | 95.8%（54.9-100.0） | 93.4%（60.9-99.9） |
| specificity | | | |
| Reitsma | 96.3%（93.8-97.8） | 78.5%（57.4-90.8） | 92.9%（87.2-96.2） |
| Latent | 94.3%（49.1-100.0） | 73.8%（0.9-100.0） | 93.1%（47.1-100.0） |

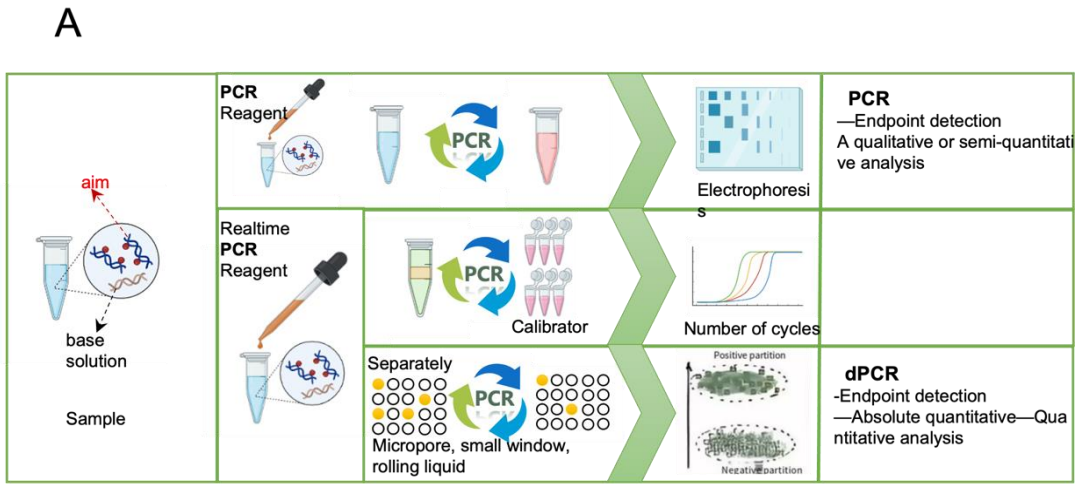

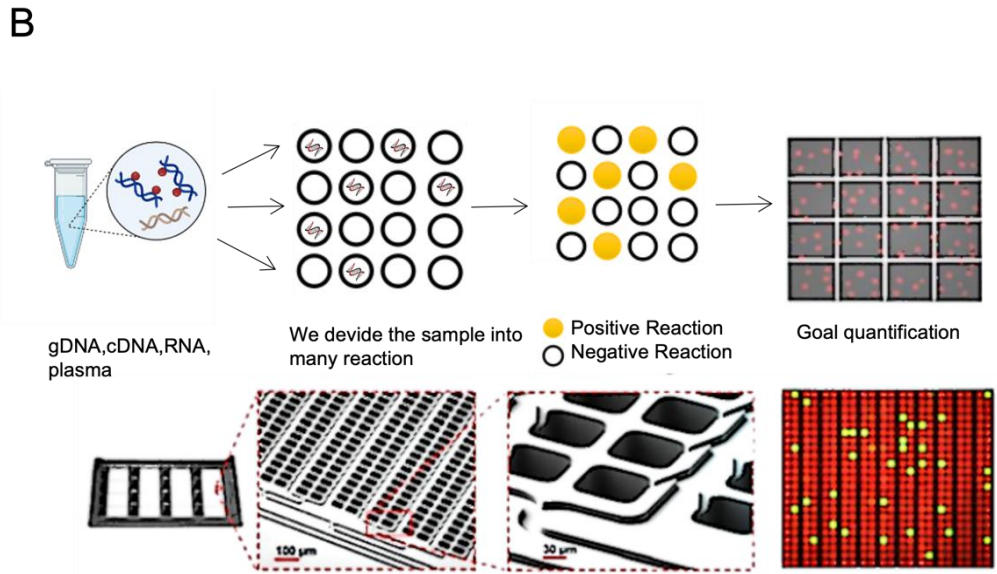

**Fig 3. The figure depicted the testing methodology of SARS-CoV2 in the wastewater base epidemiology. (A) Operation process of each generation of PCR; (B) Microreactor and chip-based analyses.**

## 6. Calculation of the number of patients infected with SARS-CoV-2 in sewage epidemiology

In the actual use of WBE, the complexity and diversity of its implementation and environment, as well as the geographical and economical constraints imposed on the regions, lead to the results of estimating the number of infections are often significantly different from the actual situation [40,65–69]. This section focuses on the data and mathematical methods of studies related to the new impact factors for calculating the infectious status in some of the WBE-related reports and papers published from 20 years to the present. The main factors are fecal shedding level of infected population, seasonal temperature, population density, and economic situation, etc. Also, the concentration of viral RNA in sewage, sewage plant treatment level, and sewage sampling sites are critical for prediction accuracy [21,70–75].

### 6.1 Effect of fecal shedding rate on the number of infections

One key finding from published studies is that SARS-CoV-2 shedding was detected in the excreta of infected individuals, among others, i.e., the virus is shed from feces in sewage [76]. This allowed the analysis of viral infectivity and the main research direction was its shedding level [77,78]. Scholars have introduced a factor for WBE, which also effectively improves accuracy. The following equation is about viral shedding combined with WBE infection rate estimation.

$$P_{WBE} = \frac{F \times C_{RNA}}{P_C \times R_S \times E} \times 10^6 \quad (1)$$

$$P_{WBE} = \frac{C_{RNA} \times Q_W}{R_S \times E} \times 10^5 \quad (2)$$

In the equation, $P_{WBE}$ is the COVID−19 infection rate per 100,000 people in the test area, F is the daily sewage flow (megaliters/day); $P_c$ is the population of the community (100,000 people as a unit); $R_s$ is the fouling index calculated for SARS−CoV−2 in sewage due to decay and other effects, related to sewer acidity and other conditions and shedding levels, and E is the daily amount of excreted virus (in viral gene replication/per patient per day); $Q_W$ is the average daily water consumption (liters/per person per day) [79].

The factors associated with the level of shedding, namely $R_s$ and E, were accessed to calculate the index of sewer scaling and the approximate amount of viral RNA excreted by one patient per day.

Viral shedding also occurs not only in patients carrying the virus [80], and studies have shown that shedding may continue for up to a month even when patients are cured [81–83]. In a sampling report, a follow-up analysis of 2149 patients from different cities in several countries, including after recovery, revealed a positive rate of 51.8% in their stool samples [84].

### 6.2 Effect of temperature and virus decay rate

Studies and news in the last two years have shown that the higher the temperature, the faster the virus decays, and in populated areas such as communities, the hotter the weather, the slower the news of infection seems to spread, while it so happens that outbreaks of new crowns, so far tend to be large-scale outbreaks in cooler weather [85,86]. Some experiments have shown $T_{90}$ values for RNA of SARS-CoV-2 ranging from 8.04 to 27.8 days in sewage and 9.4 to 58.6 days in tap water at 37 °C and 4 °C, respectively [87]. There were experimental attempts to describe the decay pattern using log-linear and nonlinear models, and nonlinear models were obtained by nonlinear least squares, resulting in a $T_{90}$ of 1.2 days for SARS−CoV−2 in pasteurized unfiltered effluent at 24 °C, with a root mean square error of 0.0224 for the Weibull model; at 4 °C, SARS− CoV−2 at 4°C with a $T_{90}$ value of 5.5 days and a root mean square error of 0.0619 for the Weibull model, in addition the authors reprocessed the data from several other experiments to obtain the root mean square error and $T_{90}$ values for the best model [88], as shown in Table 5.

Table 4: $T_{90}$ estimates for SARS-CoV-2 in different matrices (data are obtained in unfiltered effluent, where temperature superscripts with "*" are for experiments performed in pasteurized effluent)

| Virus | Temperature | Best-fit model | Root means square error | $T_{90}(d)$ | Source |
|---|---|---|---|---|---|
| SARS-CoV-2 RNA | 37 | First order log-linear model | 1.10 | 0.74 | [87] |
| | 25 | First order log-linear model | 0.67 | 12.6 | |
| | 15 | First order log-linear model | 0.59 | 20.4 | |
| | 4 | First order log-linear model | 0.37 | 27.8 | |
| | 37* | First order log-linear model | 0.59 | 5.71 | |
| | 25* | First order log-linear model | 0.48 | 13.5 | |
| | 15* | First order log-linear model | 0.32 | 29.9 | |
| | 4* | First order log-linear model | 0.14 | 43.2 | |
| SARS-CoV-2 | 20 | Log-linear model | 1.8 | 1.6 | [89] |
| | 50 | Log-linear model | 1.4 | 15 min | |
| | 70 | Log-linear model | 1.9 | 2.2 min | |

From Table 5 it can be seen that temperature has a significant effect on the decay rate of coronaviruses and that temperature may alter the environmental half-life of indicators of virus transmission in effluent by 27% to 7,010% [90]. One study calculated first-order decay rate constants and associated 95% confidence intervals for SARS-CoV-2 at specific temperatures:

$$\ln\left(\frac{C_t}{C_o}\right) = -k \times t \quad (3)$$

$$T_{90} = -\frac{\ln(0.1)}{k} \quad (4)$$

$C_t$ and $C_0$ are the viral RNA concentration (GC/mL) at moment t and moment 0, respectively, and k is the decay rate constant. Use Equation (4) to calculate the time required to reach 90% (a logarithmic) reduction ($T_{90}$).

Temperature was the most important environmental variable contributing to the observed changes in the dataset (>80%), followed by the state of the water (~5%) [87]. While comparing the single-phase and two-phase attenuation models with additional square and F-tests revealed that the two-phase attenuation model did not improve the fit to the observed SARS-CoV-2 data in effluent and water ( p > 0.05) compared to the single-phase attenuation model (Equation 1), therefore, using the single-phase attenuation model may be the superior approach [89].

It has been suggested that the degradation of biomarkers present in the effluent over time can be expected to follow an exponential decay [91], described by the equation:

$$t_{\frac{1}{2},2} = t_{\frac{1}{2},1} \times \frac{\ln 2}{\ln 2 \times Q_{10}^{\frac{T_1-T_2}{10°C}}} \quad (5)$$

Here $t_{\frac{1}{2},1}$ is the initial half–life, $T_1$ is the temperature at which the initial half-life was derived, $t_{\frac{1}{2},2}$ is the half–life at seasonally and spatially adjusted sewer temperatures calculated in this study, $T_2$ is the calculated temperature to which the initial half–life was adjusted, and $Q_{10}$ is a temperature-dependent factor for rate variation ranging from 2 to 3 for most biological systems [34]. Data from another study also partly demonstrate that the spread of the virus is affected by temperature[92–98], and Figure 4 (A) shows the timing and number of infections in the global and global regions, with the number of patients troughing during the warmer summer months[99]. Figure 4(B) shows the relationship between average monthly minimum ambient temperatures (AMMAT) and ratio of infection incidence in mammals and human (Pos/Tot), which was determined based on coronavirus detection by PCR. Red line is a linear fitting with minimum root mean square deviation (RMSE). Blue tape describes the range of cumulative infected ratio in top 31 countries worldwide[100]. Figure 4 (C-E) presents data on how temperature affects the ability of SARS-CoV-2 to cause infection on non-living surfaces. In Figure 4 (C), measurements of $TCID_{50}$/mL values of recovered virus are shown at different time points in hours, at 4°C (blue panel), room temperature (green panel), and 30°C (red panel). The dots represent the average of three independent experiments, and the standard error is shown as bars. Dashed lines indicate the lower limit of quantification. In Figure 4 (D), regression plots are used to show the predicted decay of the virus over time, with each dot representing the $TCID_{50}$/mL value of a single experiment. Fifty lines per replicate indicate possible decay patterns for each experimental condition. Finally, Figure 4 (E) displays the estimated half-life ranges of the virus at different temperatures using violin plots. The dots indicate the median estimates with 95% confidence intervals[101].

## 7. Prospects and Challenges

Wastewater-based epidemiology allows for the rapid acquisition of infection information in surrounding areas of collection points. The use of machine learning or deep learning algorithms can predict changes in COVID-19 infection numbers. By processing influencing factors such as environmental temperature and biological markers, more timely and accurate predicted results are expected. This will help advance trend prediction of the number of infected individuals, risk warning, and assessment of lockdown effectiveness, among other tasks. It will also greatly promote local government and medical institutions to allocate epidemic prevention resources in advance and carry out scientific epidemic prevention work. Due to the relatively low cost of wastewater-based epidemiology testing, it can effectively reduce the financial burden of local COVID-19 testing.

However, due to the specificity of the sewer environment and the limited number of SARS-CoV-2 experiments based on wastewater-based epidemiology, there are still many uncertain factors that interfere with predictions, such as pH value of sewage, precise measurement of sewer temperature, virus residence time in the sewer, and so on. The conclusions of some studies are conflicting or controversial, but this may also be due to differences in geographic environment among experimental sites or countries, suggesting that a universal and accurate prediction method or model for SARS-CoV-2 based on WBE has not yet been found. These unknown and uncertain influencing factors will also be key research directions in the future.

**Author Contributions:**

Conceptualization, F.J.H., Y.Z.C.; methodology, C.H.Q.D., F.J.H., Y.Z.C.; investigation, Z.H.C., W.Y., Z.Z.Y., Z.X.L.,S.Y.W., P.J., J.K.F.C.; resources, M.F. and F.J.H.; data curation, M.F. and F.J.H.; writing—original draft preparation, Y.S.L., W.Y., Z.Z.Y., Z.X.L., S.Y.W., C.C., W.H.Z., T.W.; writing—review and editing, M.F., A.C.H., W.Y., J.R.L., Z.Z.Y., Y.M., Y.B.L., M.J.Y.; drawing, M.F. Z.H.C., X.X.F., Q.R.Y., S.Y.W.; supervision, C.H.Q.D.; project administration, F.J.H., Y.Z.C.; funding acquisition, F.J.H. All authors have read and agreed to the published version of the manuscript.

**Conflicts of Interest:** The authors declared no conflict of interest.

**Acknowledgements**

The authors acknowledge funding from the 2023 Shenzhen Science and Technology Innovation Committee Funds (Stable Support General Project 20231127194506001) and the 2023 Clinical Research Funds of Shenzhen Second People's Hospital (2023xgyj3357009).